\documentclass[12pt]{article}
\usepackage{epsfig}
\usepackage{latexsym}
\usepackage{amssymb}
\usepackage{amsfonts}

\usepackage{axodraw}
\usepackage{color}

\newcommand{\be}{\begin{equation}}
\newcommand{\ee}{\end{equation}}
\newcommand{\bea}{\begin{eqnarray}}
\newcommand{\eea}{\end{eqnarray}}
\newcommand{\al}{\alpha}
\newcommand{\ep}{\varepsilon}
\newcommand{\sg}{\sigma}
\newcommand{\pa}{\partial}
\newcommand{\dd}{\mbox{d}}

\begin{document}
\parindent=1.5pc

\begin{titlepage}
\rightline{\footnotesize TTP-Number: TTP08-30}
\rightline{\footnotesize SFB-Number: SFB/CPP-08-51}

\vspace{2cm}

\bigskip
\begin{center}
{{\large\bf Algorithm FIRE --- Feynman Integral REduction} \\
\vglue 5pt \vglue 1.0cm {\large A.V. Smirnov}\footnote{
Supported by RFBR grant 08-02-01451-a and DFG through SBF/TR 9.
}\\
\baselineskip=14pt \vspace{2mm} {\normalsize Scientific Research
Computing Center of Moscow State University
\\
Institut f\"{u}r Theoretische Teilchenphysik - Universit\"{a}t Karlsruhe
} \baselineskip=14pt
\vspace{2mm} \vglue 0.8cm {Abstract}}
\end{center} 
\vglue 0.3cm {\rightskip=3pc
 \leftskip=3pc
\noindent The recently developed algorithm FIRE performs the
reduction of Feynman integrals to master integrals. It is based on
a number of strategies, such as applying the Laporta algorithm, the $s$-bases
algorithm, region-bases and integrating explicitly over loop momenta when possible.
Currently it is being used in complicated three-loop calculations.
 \vglue 0.8cm}
\end{titlepage}


\section{Introduction}

In the framework of perturbation theory quantum field theoretical
amplitudes are written as sums of Feynman integrals that
are constructed according to Feynman rules.
After a tensor reduction 
each Feynman graph generates various scalar Feynman integrals
with the same structure of the integrand but with various
powers of propagators (also called {\em indices}):
\bea
  F(a_1,\ldots,a_n) &=&
  \int \cdots \int \frac{\dd^d k_1\ldots \dd^d k_h}
  {E_1^{a_1}\ldots E_n^{a_n}}\,.
  \label{eqbn}
\eea
Here $k_i$, $i=1,\ldots,h$, are loop momenta
and the denominators $E_r$ are either quadratic or linear with respect
to the loop momenta $p_i=k_i, \; i=1,\ldots,h$ or
to the independent external momenta $p_{h+1}=q_1,\ldots,p_{h+N}=q_N$ of the graph.
Irreducible polynomials in the numerator can be represented as
denominators raised to negative powers.
Usual prescriptions $k^2=k^2+i 0$, etc. are implied
and dimensional regularization \cite{dimreg} with $d=4-2\ep$ is assumed.

In today's perturbative calculations, when one needs
to evaluate millions of Feynman integrals (\ref{eqbn}), a
well-known strategy is to derive certain relations between Feynman
integrals of a given family without calculating them, and
then to apply the latter {\it recurrently} in order to find an
algorithm that expresses any given Feynman integral as a linear
combination of some {\it master} integrals.

There are several types of commonly used relations, but the most
important are the the so-called integration by parts (IBP)
relations \cite{IBP}
\bea \int\ldots\int \dd^d k_1 \dd^d k_2\ldots
\frac{\pa}{\pa k_i}\left( p_j \frac{1}{E_1^{a_1}\ldots E_n^{a_n}}
\right)   =0  \;. \label{RR}
\eea
The derivatives of the scalar products $k_i\cdot k_j$ and $k_i\cdot q_j$ can be expressed
linearly in terms of the factors $E_i$ of the denominator, hence one
obtains the IBP relations in the following form:
\begin{equation}
\sum \al_i F(a_1+b_{i,1},\ldots,a_n+b_{i,n})
=0\,,
\label{IBP}
\end{equation}
where $b_{i,j}$ are some fixed integers and $\al_i$ are polynomials in $a_j$.
Now one can substitute all possible $(a_1,a_2,\ldots,a_n)$ on the left-hand sides
of (\ref{IBP}) and obtain a large number of relations.

There are several recent attempts to make the reduction procedure
systematic, in particular, the so-called Laporta algorithm
\cite{LGR1,LGR2} (There is a public version named AIR which implements the
algorithm on a computer \cite{AnLa}) and Baikov's
method (see \cite{Bai,ST} and chapter 6 of \cite{EFI}). Another activity in this direction is connected
to the use of Gr\"obner bases \cite{Buch}. The first variant of
this approach was suggested in \cite{Tar1}, where IBP relations
were reduced to differential equations. First attempts to use
directly the non-commutative Gr\"obner bases in the algebra
generated by shift operators were made in \cite{Gerdt,GeRo}.

We presented another approach based on Gr\"obner bases in
\cite{ourwork}. More information can be found in \cite{ourwork2}
and the algorithm constructing the $s$-bases has been described in
\cite{mypaper}. This algorithm called {\tt SBases} is now made public
and is available at http://www-ttp.particle.uni-karlsruhe.de/$\sim$asmirnov/.
This paper provides the syntax to use it, however its internal
structure will not be discussed here.

This paper describes the algorithm FIRE, combining a number of
ideas from different existing algorithms. One of the important
parts of the algorithm is the use of $s$-bases
(the original idea of FIRE three years ago was to construct the $s$-bases in all sectors before the reduction). However,
FIRE can also work in a ``pure Laporta mode'', i.e. without using
the $s$-bases.
To run FIRE in this mode one simply requires to provide
the IBPs, symmetries and boundary conditions.
However, in practice this part of FIRE is applied only
to a small subset of integrals --- the areas
where most of the indices are positive are usually done with the
$s$-bases part and when there are enough negative indices, an
explicit integration can be performed.

Although the paper is devoted to the algorithm FIRE,  one also requires
some other codes that are not a part of FIRE. All of them are available at
http://www- ttp.particle.uni-karlsruhe.de/$\sim$asmirnov/, and the needed instructions
for their usage are provided. However, those projects will not be discussed
in detail. Both FIRE and those codes are distributed under the terms of GNU GPLv2.

FIRE can be downloaded directly from
http://www-ttp.particle.uni-karlsruhe.de/ $\sim$asmirnov/data/FIRE\_3.0.0.m
and a package with FIRE and supplementary codes and binaries is available via
http://www-ttp.particle.uni-karlsruhe.de/ $\sim$asmirnov/ data/FIRE.tar.gz.

To use all features of FIRE one requires the following codes:
{\tt FIRE\_3.0.0.m}\footnote
{The first version was a private one, while the second version was shared with collaborators.} --- the algorithm FIRE itself, written in Wolfram Mathematica;
{\tt IBP.m} --- a small code for creating IBPs;
the {\tt SBases\_3.0.0.m} --- the code to construct the $s$-bases.

To take most out of the code, one should also use the QLink package
(the package also has its own page, http://qlink08.sourceforge.net/) that is used to
access the QDBM database for storing data on disk from Mathematica;
and the small FLink tool that allows
to perform external evaluations by means of the Fermat program.
The usage of the code and the related packages will be discussed in section 5.

Let us start with explaining the idea of FIRE.
The first step will consist in introducing an ordering
on Feynman integrals.

\section{The ordering}

As it has been noted in \cite{ourwork2}, to
define master integrals, or irreducible integrals, we need to
choose a certain priority between the points $(a_1,\ldots,a_n)$,
formally to introduce a complete ordering on them. Moreover, we
are intending to consider integrals corresponding to different Feynman
diagrams (to use integration over some loop momenta), therefore
we actually need a global ordering on pairs
$(DN,(a_1,\ldots,a_{n(DN)}))$, where $DN$ is the diagram number.

We introduce such an ordering in three steps. First of all we
enumerate Feynman diagrams and say that all
integrals corresponding to a diagram with a bigger number are
lower than the ones corresponding to a diagram with a smaller one
 (the ordering will be
denoted with the symbol $\prec$ and named as \textit{lower}).

Now let us fix a Feynman diagram and introduce an ordering
on integrals corresponding to it. There are
different ways to do that, but surely one is going to choose
something natural, so that the integrals corresponding to the
minimal elements in this ordering are easier to be calculated.

Let us
realize that the Feynman integrals are simpler, from the analytic
point of view, if they have more non-positive indices. In fact, in
numerous examples of solving of IBP relations by hand, the natural
goal was to reduce indices to zero or negative values. Moreover,
the reduction procedure is often stopped once one arrives at
integrals which are already sufficiently simple, in particular,
when they can be expressed analytically in terms of gamma
functions for general dimension $d$. Experience reflected in
many papers leads to the natural idea to decompose the whole
region of the integer indices which we call {\em
sectors}\footnote{ This decomposition is also standard within
Laporta's algorithm \cite{LGR1,LGR2}. Here one applies the word
{\em topology} which we tend to avoid because this is an
example of a situation where a commonly accepted mathematical term
is used as a slang word to denote something else.}.

Let us consider the set $\cal D$ with elements
$\{d_1,d_2,\ldots,d_n\}$ where all $d_i$ are equal to $1$ or $-1$.
The elements of this set are called \textit{directions}. For any
direction $\nu=\{d_1,\ldots,d_n\}$ we consider the region $\sg_\nu
= \{ (a_1,\ldots,a_n): (a_i - 1/2)  d_i > 0\}$ and call it {\em a
sector}. In other words, in a given sector, the indices
corresponding to $1$ are positive and the ones corresponding to $-1$ are non-positive. Obviously
the union of all sectors contains all integer points in the
$n$-dimensional vector space and the intersection of any two
sectors is an empty set. We will say that a direction
$\{d_1,\ldots d_n\}$ is lower than $\{d'_1,\ldots d'_n\}$ if
$d_1\leq d'_1,\ldots,d_n\leq d'_n$ and they are not equal
\footnote{This is not a complete ordering on directions.
In practice one has to be able to compare all directions,
but the choice does not influence the reduction procedure much.}. The
analogue definition is also applied to the corresponding sectors.

Thus, the second natural step is to define that $F(a_1,\ldots,a_n)\succ
F(a'_1,\ldots,a'_n)$ if the sector of $(a'_1,\ldots,a'_n)$ is
lower than the sector of $(a_1,\ldots,a_n)$
\footnote{This is a certain simplification of the ordering used in FIRE,
to introduce it completely we have to introduce the notion of regions
and the way to compare them, that will be done in \ref{regions}.}.

To define an ordering completely we introduce it in some way
inside the sectors. Let us fix a sector $\sg_\nu$. Any point in
this sector can be written as
$(a_1,\ldots,a_n)=(p_1+d_1 b_1,\ldots,p_n+d_n b_n)$,
where $(p_1,\ldots,p_n)=((d_1+1)/2,\ldots,(d_n+1)/2)$ is the
\textit{corner} point of the sector, and all $b_i$ are non-negative.
This sets a one-to-one correspondence $\phi$ between points $(a_1,\ldots,a_n)$ of
the sector $\sg_\nu$ and
$(b_1,\ldots,b_n)\in\mathbb N^n$.
The latter set is a semi-group
(with respect to $(b_1,\ldots,b_n)+(b'_1,\ldots,b'_n)=(b_1+b'_1,\ldots,b_n+b'_n)$).
We will say that an ordering on $\mathbb N^n$ is
\textit{linear} if
\\
i) for any $a\in \mathbb N^n$ not equal to $(0,\ldots 0)$ one has $a\succ (0,\ldots 0)$
\\
ii) for any $a,b,c\in \mathbb N^n$ one has $a\succ b$
if and only if $a+c\succ b+c$.

An ordering on $\sg_\nu$ will be also named linear if the
corresponding ordering on $\mathbb N^n$ is linear
(we say that $\phi(x)\succ\phi(y)$ if and only if $x\succ y$).
Now the third step is to fix a linear ordering in every sector
\footnote{All known Laporta algorithms also use the linear orderings.
Moreover, the set of linear orderings is big enough,
and only a small subset of those is used in practice.}.

\section{The algorithm}

The idea of FIRE is that at every moment we store the
so-called \textit{proper expressions} of Feynman integrals,
meaning that the integral $F(a_1,\ldots,a_n)$ is expressed as a linear combination of
integrals that are \textit{lower} than $F(a_1,\ldots,a_n)$.

There are different ways of producing proper expressions for
Feynman integrals, that we will discuss in the next section.
Suppose now that there is a way to obtain a proper expression for
any integral. Then the main algorithm looks this way
(where Input is the set of integrals one wants to calculate):

\vspace{1cm}

\textit{FIRE}
\begin{tabbing}
1.    \=$RequiredIntegrals$={Input}
\\
2.\>\textbf{While} \= $RequiredIntegrals$ is non-empty
\\
\>\>3.  \= $Y$ = $RequiredIntegrals[[1]]$
\\
\>\>4.\> Remove the first element of $RequiredIntegrals$
\\
\>\>5.\> \textbf{Obtain a proper expression for $Y$}
as a linear combination
\\
\>\>\>of integrals; we will denote this set with $S$
\\
\>\>6.\> Take the subset $S'$ of $S$ containing integrals
\\
\>\>\>that do not yet have proper expressions
\\
\>\>7.\> Unite $RequiredIntegrals$ with $S'$
\\
8.\>\textbf{EndOfWhile}
\\
9.\>Sort the list of all integrals encountered so that
\\
\>\>
the higher
ones are at the beginning of the list.
\\
10.\>\textbf{For} all elements $Y$ of the list, starting from the
end
\\
\>\>11. Take the existing proper expression of $Y$ by $(X_1,\ldots,X_k)$
\\
\>\>\>and substitute into it the proper expressions of $X_i$
\\
12.\>\textbf{EndOfFor}
\\
13.\>Return the obtained expression for Input
\end{tabbing}

In other words, we start from an integral, obtain a proper
expression for it, look at the integrals it was expressed by and
so on. Since every integral is expressed in terms of lower
integrals, this procedure will stop at some point.
Then it will be left to sort the list and substitute everything
backwards. It is worth noting that the length of expressions does
not grow on the substitutions cycle. Indeed, such a length is
limited by the number of irreducible integrals.

But all this was based on being capable of finding proper
expressions for integrals (step 5), preferably doing this fast. If one
could construct those by a constant amount of time, then
one could expect to reduce any needed integral to master integrals.
This can be done by the use of $s$-bases. However,
currently one can not construct those for all sectors, so a
combination of methods is needed.

\section{Proper expressions}

\label{proper}
There are different sources of proper expressions. The power of
FIRE is in using both traditional sources and at the same time the
$s$-bases technique.

The most commonly used source consists
of applying \textit{the Laporta algorithm} (see \cite{LGR1,LGR2}) to derive
proper expressions from IBP relations \cite{IBP}.
The basic idea of this algorithm is to fix a sector and
start generating IBP relations with different index substitutions.
Since the number of relations grows faster that the number of
integrals, at some point the linear system becomes overdetermined
and can be solved.

It is important to note that if the Laporta algorithm is applied
in a sector, then we perform the so-called \textit{tail-masking} \cite{AnLa} meaning
that if an IBP also includes some integrals from lower sectors
then that part is masked and is not substituted anywhere.
Otherwise, the growth of those parts would not let the algorithm
work.

Second, one can use the \textit{symmetries} of the diagram.
Typically they have the following form:
\begin{equation}
F(a_1,\ldots,a_n)=(-1)^{d_1 a_1+\ldots d_n
a_n}F(a_{\sigma(1)},\ldots,a_{\sigma(n)}),
\label{Sym}
\end{equation}
where $d_i$ are fixed and are equal to either one or zero, and $\sigma$ is a
permutation.

Next, one can use \textit{boundary conditions},
i.e. the conditions of the following form:
\be
F(a_1,a_2,\ldots,a_n)=0\mbox{ when }a_{i_1}\leq 0,\ldots, a_{i_k}\leq 0
\label{boundary}
\ee
for some subset of indices $i_j$. (In particular, we always have
$F(a_1,a_2,\ldots,a_n)=0$, if all $a_i$ are non-positive).

Furthermore, there are the so-called \textit{parity conditions},
stating that Feynman integrals should
be zero if the sum of some subset of indices is odd.

One more source of proper expressions is the usage of \textit{manually-inserted
rules}. For example, one might consider several similar diagrams at the same time
(which happens quite often in practice) so that if some indices turn negative,
the diagrams become equal. This relation is provided into the code
by inserting the rules manually. One more important usage of rules
is related to the region-bases and will be discussed later.

Now, what can be called a bridge between Laporta and $s$-bases
approaches, is the \textit{direct use of IBP relations}.
There are too many IBP relations and surely there comes the
question of using them better. One of the ideas is to look at a
given integral and try to pick an IBP such that after substituting
certain indices this integral $X$ is the highest one among the ones in
the relation. In this case one would immediately obtain a proper
expression for $X$.

However, this is not applicable to all integrals.
But still, one may consider a sector and
reduce the integrals while possible with this method.
Afterwards one can apply the Laporta algorithm
in this sector, however the number of IBPs to be generated
becomes smaller.

The direct usage of IBPs can been replaced in the latest version of FIRE
by the usage of ideas of Lee \cite{Lee}.
Basically, in each sector one may find a single IBP that generates
proper expressions for ``most'' points in this sector.
Accordingly, we might generate less IBPs because the other IBPs are naturally
represented as a linear combination of the ones we generated.
For details see Criterion 3 in \cite{Lee}.


FIRE would loose much of its functionality if it did not use
the $s$-bases approach to the reduction problem.
One can say that an $s$-basis in a sector is a mechanism of obtaining a proper
expression for any (except for master integrals) integral in this
sector by a constant\footnote{
Speaking formally, for each $s$-basis we can
specify the maximum number of numerical operations
required to construct a proper expression with the use of this basis.} amount of time.

The idea of $s$-bases comes from the above-mentioned direct usage
of IBPs. As it has been said, the initial relations cannot be
applied to construct proper expressions directly for every
integral. Therefore one
might try to find a better set of relations $S$, such that for any
integral $X$ (except for master integrals) one might find an element
of $S$ and substitute some indices to obtain a proper expression
for $X$.

To do this one first needs a language to work with
relations before substituting the indices. And that is the
language of algebras of operators acting on the field of
functions. The algorithm aimed at constructing the $s$-bases is
close to the Buchberger algorithm. However, it does not always
succeed, and it is a skill to find a good ordering for it to work.
The details on the construction of $s$-bases can be found in \cite{mypaper}.

Finally, let us explain \textit{the usage of region-bases}.
Let us expand the set of directions $\cal D$
up to the sets of numbers
$\{r_1,r_2,\ldots,r_n\}$ where all $r_i$ are equal to $1$, $-1$ or $0$.
For any $\nu=\{r_1,\ldots,r_n\}$ we consider the set $\sg_\nu
= \{ (a_1,\ldots,a_n): (a_i - 1/2)  r_i \geq 0\}$ and call it {\em a
region}. In other words, in a given region, the indices
corresponding to $1$ are positive, the ones corresponding to $-1$ are non-positive
and the ones corresponding to $0$ are arbitrary. With this definition we can turn back to the
algorithm description.

Experience shows that $s$-bases are constructed easily if the number of
non-positive indices in a given sector is small.
And if the number of non-positive indices is large, there is usually a
possibility to perform integration over some loop
momentum explicitly in terms of gamma functions for general $d$.
However,
to derive the corresponding explicit formulae, with multiple
finite summations, for all necessary cases turns out to be
impractical.

In this situation, there is another alternative. Let us
distinguish the propagators (and irreducible numerators) involved
in such an explicit integration formula. They are associated with a
subgraph $\Gamma'$ of the given graph $\Gamma$. Let us solve the IBP relations for the
corresponding subintegral in order to express any such subintegral
in terms of some master integrals.
The coefficients in these reduction formulae depend not only on $d$
and given external parameters but also on the kinematic invariants of $\Gamma$
which are external for $\Gamma'$.

If the highest coefficients of the $s$-basis are monomials
of those kinematic invariants, the reduction
for the $\Gamma'$ can be ``raised'' to the diagram $\Gamma$.
It turns into a reduction in a region, where the indices corresponding
to $\Gamma\backslash\Gamma'$ are arbitrary\footnote{
The usage of regions obviously
affects the definition of the ordering.
See section \ref{regions} for more details.}.

After using this reduction
procedure, it is sufficient to use explicit integration
formulae only for some boundary values of the indices
(corresponding to master integrals of $\Gamma'$). This
replacement is very simple, without multiple summations.

Integrals which are obtained from initial integrals by an explicit
integration over a loop momentum in terms of gamma functions
usually involve a propagator with a regularization by an amount
proportional to $\epsilon$. FIRE is also applicable in such
situations.

In some cases one cannot perform an explicit integration
because the diagram contains a massive-massive or a massive-massless bubble.
Still one can reduce the indices on those lines to $(1,1)$ or $(1,0)$
by means of a region basis. Afterwards the diagram might
be mapped to a new diagram with one loop less and containing a so-called
\textit{heavy point} corresponding to this bubble.
The {\tt HeavyPoints} setting will instruct the code that there is
an index whose value can be only $0$ and $1$. However,
the IBPs for such a diagram should be constructed manually.

Now what FIRE actually does when it needs to construct a proper
expression for a given integral, is
\begin{itemize}
\item checking if it is equal to zero by boundary conditions;
\item checking if it is equal to zero by parity conditions;
\item looking at its symmetry orbit checking if it is not minimal there;
\item looking for manual rules to map this integral somewhere;
\item constructing a proper expression using an $s$-basis if there is one in the corresponding sector (or a region basis if this point lies in a region);
\end{itemize}

Surely, if FIRE manages to construct a proper expression for the
integral by one of the methods, it does not try the other ones for
this integral but moves to other integrals instead.

For the remaining integrals FIRE launches the Laporta algorithm in the
highest sector. It results in a number of new proper expressions
and lower integrals that are to be reduced, and the procedure is repeated.

\section{Benchmarks}

Up to the moment of the publication there is only one public Laporta algorithm available,
AIR \cite{AnLa} by Anastasiou and Lazopoulos. Although FIRE is not only a Laporta algorithm,
we had to compare the algorithms in different modes. To avoid problems related to peculiarities
of algorithms, we used a very simple example --- the massless on-shell box diagram
also used as an example in \cite{AnLa}.

\begin{figure}[ht]
\begin{center}
\fcolorbox{white}{white}{
  \begin{picture}(135,125) (45,-5)
    \SetWidth{0.5}
    \SetColor{Black}
    \ArrowLine(75,95)(75,20)
    \ArrowLine(75,20)(150,20)
    \ArrowLine(150,20)(150,95)
    \ArrowLine(150,95)(75,95)
    \Vertex(75,20){1.41}
    \Vertex(75,95){1.41}
    \Vertex(150,95){1.41}
    \Vertex(150,20){1.41}
    \ArrowLine(165,95)(150,95)
    \ArrowLine(165,20)(150,20)
    \ArrowLine(60,95)(75,95)
    \ArrowLine(60,20)(75,20)
    \Text(150,15)[lt]{\small{\Black{$\;p_4$}}}
    \Text(75,15)[rt]{\small{\Black{$p_2\;$}}}
    \Text(112,15)[ct]{\small{\Black{$3$}}}
    \Text(75,57)[rc]{\small{\Black{$2\;$}}}
    \Text(150,57)[lc]{\small{\Black{$\;4$}}}
    \Text(150,100)[lb]{\small{\Black{$\;p_3$}}}
    \Text(75,100)[rb]{\small{\Black{$p_1\;$}}}
    \Text(112,100)[cb]{\small{\Black{$1$}}}
  \end{picture}
}
\end{center}
\caption{
\label{box}
The massless on-shell box diagram.
}
\end{figure}
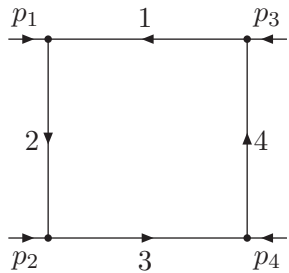

We cannot guarantee that we have found the optimal settings for AIR, however
we did a good effort to try different settings and followed the instructions
from \cite{AnLa}. The following table has the timing
for AIR running in four modes (those modes are not any internal modes of the algorithm,
but the sets of settings we used). AIR1 stands for the basic setting where one provides only the IBPs
and boundary conditions. AIR2 uses the masking of expressions: to use it one has to
provide the list of master integrals. AIR3 uses the masking of integral coefficients.
In this mode AIR does not simplify large coefficients during the reduction,
but substitutes preset numerical values to check whether they are zero. However,
coefficients have to be simplified after the reduction; we included this routine
in the time measurement. AIR4 combines both strategies.

We used FIRE also in four modes. FIRE1 is a pure Laporta mode. FIRE2 has two options turned
on --- the {\tt LeeIdeas} and {\tt DirectIBP}. FIRE3 uses the symmetries of the diagram.
FIRE4 uses pre-constructed $s$-bases.

\vspace{1cm}

\begin{tabular}{|c|c|c|c|c|c|c|c|c|}
  \hline
      &    AIR1  &     AIR2  &   AIR3   &   AIR4   &    FIRE1    &   FIRE2    &   FIRE3    &  FIRE4    \\
  10  &      28  &        8  &     14   &      9   &       14    &      15    &       6    &      3    \\
  20  &     227  &       44  &     79   &     47   &       79    &      35    &      19    &      4    \\
  30  &    1068  &      166  &    285   &    202   &      131    &      80    &      42    &      8    \\
  40  &    3268  &      567  &   1098   &    624   &      259    &     166    &      88    &     16    \\
  50  &   14561  &     2117  &   2977   &   2280   &      672    &     478    &     254    &     34    \\
  \hline
\end{tabular}

\vspace{1cm}

The setup for both algorithms and the logs can be downloaded from
http://www- ttp.particle.uni-karlsruhe.de/$\sim$asmirnov/data/FIRE-benchmarks.tar.gz

Although one cannot expect to construct the $s$-bases everywhere,
it is always possible to achieve results much better than the
pure Laporta mode by performing some work before launching the evaluating:
by constructing $s$-bases, region-bases, providing parity conditions,
providing symmetries or inserting rules manually.
Moreover, the tests of this type are unnatural for FIRE since in practice
one does not require to reduce all integrals satisfying a given condition,
but a much smaller set; FIRE is optimized for involving less integrals
if possible (especially with the use of $s$-bases).

To illustrate the potential of FIRE let us describe a problem
where FIRE was applied. It was used to calculate the fermionic contribution
to the three-loop static quark potential \cite{tlp}.
The Feynman diagrams corresponding to this problem had $12$ indices each
($11$ propagators and $1$ irreducible numerator).
The reduction method included the automatic construction
of the $s$-bases, but a number of $s$-bases was constructed
manually (by means of trying different orderings). The region bases were used as well --- they allowed
to reduce many cases to 2-loop problems (including
those with regularized lines) with $7$ indices.
The set of diagrams used during the reduction
included $18$ 3-loop diagrams and $16$ 2-loop diagrams\footnote{
We count only the so-called ``master-diagrams''.}.
The data file containing all the information on the diagrams
in an internal format had the size of $95$ MB.
We had to reduce about $70$ thousand integrals, and the task
was split into $12$ jobs. Those jobs
required maximally $2$ days, $10$ GB of RAM and $15$GB on hard disk
(with {\tt DatabaseUsage=2}, see section \ref{syntax}).
The reduction included totally about $30$ million integrals involved.

\section{Instructions}

Let us now explain how to use the different computer programs.

\subsection{IBP}

The file IBP.m is written in Mathematica and is used to create IBPs.
To use it, load the file by {\tt Get["IBP.m"]}.
The following syntax is available:

\begin{itemize}

\item \textit{Syntax:} {\tt PrepareIBP[]};

\textit{Description:} solves
the system of linear equations obtained after differentiating.
But first one has to give values to the following variables:

\begin{itemize}

\item \textit{Syntax:} {\tt Internal};

\textit{Description:} lists the loop momenta;

\textit{Example:} {\tt Internal=\{k,l\}};

\item \textit{Syntax:} {\tt External};

\textit{Description:} lists the external momenta;

\textit{Example:} {\tt External=\{\}};

\item \textit{Syntax:} {\tt Propagators};

\textit{Description:} lists the propagators and irreducible numerators;

\textit{Example:} {\tt Propagators=\{k$^2$-mm,l$^2$-mm,(k-l)$^2$\}};

\end{itemize}

\item \textit{Syntax:} {\tt IBP[x,y]}

\textit{Description:} produces an IBP obtained after multiplying
by {\tt y} and differentiating by {\tt x};

\textit{Example:} A 2-loop tadpole with two massive lines and one
massless line: {\tt Internal=\{k,l\}; External=\{\};
Propagators=\{k\^{ }2-mm,l\^{ }2-mm,(k-l)\^{ }2\}}.
\\
The IBPs
are produced by {\tt IBP[k,k], IBP[k,l], IBP[l,k] and IBP[l,l]}.

\end{itemize}

\subsection{QLink}

QLink \cite{QLink} is a package
that allows to use the QDBM database \cite{QDBM} from Mathematica. The program is
written mainly in C and uses the Mathlink technology.
FIRE can work either using QLink or not, but for heavy
calculations it becomes reasonable to use it, since otherwise
memory problems might be encountered. The knowledge of the QLink syntax
is not required; it is used directly from FIRE.

\subsection{FLink}

FLink \cite{FLink} is a tool that allows to
perform external evaluations by means of the Fermat program \cite{Fermat} which is extremely fast in working with polynomials.
FIRE can work either using FLink or not, but the usage of FLink can lead to a speed-up.

\subsection{FIRE}

\label{syntax}
{\tt FIRE\_3.0.0.m} is written in
Mathematica and is loaded with {\tt Get["FIRE\_3.0.0.m"]}.
Let us list its options and commands:

\begin{itemize}

\item \textit{Syntax:} {\tt Prepare[]};

\textit{Description:} converts the information on a diagram into the internal format;
to use this command one has to give values to the following list of variables:

\begin{itemize}

\item \textit{Syntax:} {\tt startinglist};

\textit{Description:} specifies the list of IBPs written in terms of shift
and multiplication operators. This is the same format
which is provided by the output of {\tt IBP.m}. \textit{Obligatory variable};

\textit{Example:} {\tt startinglist=\{IBP[k,k], IBP[k,l], IBP[l,k], IBP[l,l]\}};

\item \textit{Syntax:} {\tt RESTRICTIONS};

\textit{Description:} defines the boundary conditions, the list of regions
where the integrals vanish has to be specified;
it is important to specify all boundary conditions,
including symmetric ones; \textit{Obligatory variable};

\textit{Example:} {\tt RESTRICTIONS=\{\{-1,-1,0,0,0\},\{0,0,-1,-1,-1\}\}};

\item \textit{Syntax:} {\tt SYMMETRIES}, {\tt ODDSYMMETRIES} or {\tt CONDITIONALSYMMETRIES};

\textit{Description:} If all the symmetries preserve the sign, one can use the first one and
simply provide a list of possible permutations of indices (no need
to include the identical one). If the sign might be changed, one
sets the second one providing a list of pairs --- a permutation and
a set of 1s and -1s $\{s_1,\ldots,s_n\}$. In this case a point $\{a_1,\ldots,a_n\}$,
being mapped to a symmetric one, is multiplied by a product
$s_1^{a_1}\cdot\ldots \cdot s_n^{a_n}$.
A conditional symmetry is a symmetry that should be applied only in case
some if the indices have a specified sign. In this case one provides a list of pairs ---
a permutation and a region; it is important to specify all symmetries, not only the generators;

\textit{Examples:} 1) {\tt SYMMETRIES=\{\{2,1,3,4,5\},\{1,2,4,3,5\},\{2,1,4,3,5\}\}}
\\ 2) {\tt ODDSYMMETRIES=\{\{\{2,1,3,4,5\},\{1,1,1,1,-1\}\}\}}
\\ 3) {\tt CONDITIONALSYMMETRIES=\{\{\{2,1,3,4,5\},\{0,0,0,-1,0\}\}\}};

\item \textit{Syntax:} {\tt EVENRESTRICTIONS[$list1$]=$list2$};

\textit{Description:} Specifies a parity condition; $list1$ can have numbers
from $\{-1,0,1,2\}$ and $list2$ can consist only of zeros and ones.
The meaning of such a statement is that if one takes a point $\{a_1,\ldots,a_n\}$,
where the indices corresponding to $-1$ in $list1$ are
non-positive, the ones corresponding to $1$ are positive,
the ones corresponding to $0$ are equal to zero and the
remaining ones are arbitrary, then the sum of indices
corresponding to $1$ in $list2$ should be even, otherwise the
integral is equal to zero. Setting the parity restrictions is non-obligatory;
there might be multiple parity conditions for a diagram;

\textit{Example:} {\tt EVENRESTRICTIONS[\{2,2,-1\}]=\{1,1,0\}} means that if the last index
is non-positive, then the sum of the first two should be even;

\item \textit{Syntax:} {\tt RegLine} and {\tt RegLineShift};

\textit{Description:} Instructs the code that one of the diagram lines is regularized
(corresponding to a propagator with a non-integer index).
The standard shift is assumed to be $(4 - d)/2$,
but one can use any other value by setting the {\tt RegLineShift}
variable;

\textit{Example:} {\tt RegLine=5; RegLineShift=4-d};

\item \textit{Syntax:} {\tt HeavyPoints};

\textit{Description:} The {\tt HeavyPoints} option, if set, should list the indices corresponding
to heavy points; see Section \ref{proper} for details;

\textit{Example:} {\tt HeavyPoints=\{3\}};

\end{itemize}

\textit{Example:} see section \ref{example};

\item \textit{Syntax:} {\tt SaveStart[file\_without\_extension]}

\textit{Description:} Saves the information on a diagram on disk. Should be used
after successfully running the {\tt Prepare[]} command;

\textit{Example:} {\tt SaveStart["box"]} writes to {\tt box.start};

\item \textit{Syntax:} {\tt LoadStart[file\_without\_extension,$pn$]};

\textit{Description:} Loads the information on a diagram from a file.
A problem number $pn$ should be specified. It should be an integer from $1$ to $999$;

\textit{Example:} {\tt LoadStart["box",1]} loads from {\tt box.start};

\item \textit{Syntax:} {\tt LoadSBases[file\_without\_extension,$pn$]};

\textit{Description:} Loads an $s$-bases file.
Those files are produced by the {\tt SBases} code (see section \ref{SBases})
and contain the same information as the start files plus a set
of $s$-bases constructed for some sectors or regions;

\textit{Example:} {\tt LoadSBases["box",1]} loads from {\tt box.sbases};

\item \textit{Syntax:} {\tt RULES[$pn$,$region$]:=G[$pn$,$x\_$]:>...};

\textit{Description:} The rules should be prescribed after loading a
start file or an $s$-bases file. A more convenient method is to put them
in a file with the same name as the start or $s$-bases file,
but ending with {\tt .rules} instead of {\tt .start} or {\tt .sbases}.
In this case the rules will be loaded automatically.
Such a rule means that if one takes a point
in a region, then the right-hand side of the rule
should be applied to this point. It can be, for example, an
expression of the form
{\tt G[$pn$,$x\_$]:>If[$x$===$y$,something,G[$pn$,$x$]}.
This will be a rule for a special value of $x$, in all other cases it
will be ignored. If all the regions encountered in the rules are sectors,
there is nothing more required to make those rules work.
Otherwise one has to specify the list of all possible regions. The syntax will be
explained in section \ref{regions};

\textit{Example:} {\tt RULES[1,\{-1,1,0\}]:=G[1,\{x1,x2,x3\}]:>G[1,\{x2,x1,x3\}]} is an alternative
way of setting a symmetry between sectors;

\item \textit{Syntax:} {\tt CreateProblem[$pn$,$n$]};

\textit{Description:} A fast way to add a diagram where you are not going
to use anything but rules. $n$ stands for the number of indices;

\textit{Example:} {\tt CreateProblem[2,5]};

\item \textit{Syntax:} {\tt Burn[]};

\textit{Description:} This command should be executed after loading
or preparing all start files, $s$-bases files and rules.
The command performs some internal optimizations and enumerations
allowing to speed-up the algorithm.  {\tt Burn[]} can work slowly at high
dimensions (especially if one uses multiple diagrams);

\item \textit{Syntax:} {\tt SaveData[file]};

\textit{Description:} Saves the internal data produces by the
{\tt Burn[]} command in a file. This file contains the information
on all diagrams and some internal data and should not be edited manually;

\textit{Example:} {\tt SaveData["all.data"]};

\item \textit{Syntax:} {\tt LoadData[file]};

\textit{Description:} Loads the data from the file. One does not
require to run the {\tt Burn[]} command after loading the data,
however no more diagrams can be added at this moment;

\textit{Example:} {\tt LoadData["all.data"]};

\item \textit{Syntax:} {\tt F[$pn$,$\{a_1,\ldots,a_n\}$]};

\textit{Description:} The main command to evaluate
the integral from diagram number $pn$ and indices $\{a_1,\ldots,a_n\}$.
The answer comes as a linear combination of terms
looking like {\tt G[$pn$,$\{a'_1,\ldots,a'_n\}$]}. This means
completely the same as $F$, but it is an internal convention, that
$G$ is something that is left unevaluated and $F$ makes the
algorithm work;

\textit{Example:} {\tt F[1,\{2,3,0,-4\}]};

\item \textit{Syntax:} {\tt SaveTables[File(obligatory), IntegalList(non-obligatory), \\SaveSymmetric(non-obligatory)]};

\textit{Description:} Saves the tables produced during the evaluation in a file.
The integral list can be
missing, in this case all tables are saved. This option is not
recommended and can result in memory overflow; and in real
problems one will need only thousands of values, while there can
be millions and more stored in tables. The {\tt SaveSymmetric}
option is assumed to be {\tt False}, but if it is {\tt True}, the
tables for symmetric integrals are also saved. This might be
useful and save time if one did a long computation of integrals,
that are not minimal in their symmetry orbits, and might need the
symmetrical integrals later;

\textit{Example:} {\tt SaveTables["1.tables",\{\{1,\{2,3,0,-4\}\},\{1,\{2,3,0,-3\}\}\}]};

\item \textit{Syntax:} {\tt LoadTables[File]} or {\tt LoadTables[FileList]};

\textit{Description:} Loads the tables from a file or a list of files.
Please keep in mind that one cannot run this
command twice without quitting the kernel or do a calculation and
then load some tables. This is done for the reason that same
integrals might have different numbers in different calculations
and it is not easy to combine them together. However, the
{\tt LoadTables[FileList]} syntax gives all the functionality one
needs. For example, if one has done an evaluation and now wants to
load some tables, one can first save the contents of the memory,
quit the kernel, then load the tables together;

\textit{Example:} {\tt LoadTables["1.tables"]};

\item \textit{Syntax:} {\tt ClearTables[]};

\textit{Description:} Clears the tables from memory.

\item \textit{Syntax:} {\tt EvaluateAndSave[ListOfIntegrals,FileForTables]};

\textit{Description:} The alternative mode ``evaluate and save'' to use FIRE.
In this mode the code performs a cleanup regularly that allows a real memory economy.
However, this cleanup makes impossible to work with the code after the evaluation
is over, so it is best to use this mode in batch jobs, where one
loads the initial data and gives a command to evaluate and save the integrals;
afterwards the job should be terminated;

\textit{Example:} {\tt EvaluateAndSave[\{\{1,\{2,3,0,-4\}\},\{1,\{2,3,0,-3\}\}\},"1.tables"]};

\item Let us now list the options that FIRE has.

\begin{itemize}
\item {\tt UsingIBP}: usually set to {\tt True}, but if switched
to {\tt False} turns the Laporta part off. Only possible
if $s$-bases are available;
\item {\tt DatabaseUsage}: a number between $0$ and $3$, determines
how heavily the QDBM database should be used;
\item {\tt QLinkPath}: should be set properly to use the database,
must point at the QLink program;
\item {\tt DataPath}: should be set properly to use the database;
the code creates up to four directories with the path starting at {\tt DataPath},
each of those being a QDBM database;
\item {\tt UsingFermat}: a setting indicating whether the Fermat
program is used for substitutions;
\item {\tt FLinkPath}: should be set properly to use Fermat,
must point at the FLink program;
\item {\tt FermatPath}: should point at the Fermat program in order to use it;
\item {\tt DirectIBP} (available in version 3.0.1): if true, the code uses IBPs directly as if it is an $s$-basis, see section \ref{proper};
\item {\tt LeeIdeas}: if true, the code omits some of the IBPs according to the ideas of Lee \cite{Lee}.
\end{itemize}

\end{itemize}

\subsection{SBases}

\label{SBases}
{\tt SBases\_3.0.0.m} is the code used to construct the $s$-bases. It is written in
Mathematica and is loaded simply with {\tt Get["SBases\_3.0.0.m"]}.
The underlying algorithms are not a subject of this paper, however, we will
provide the usage instructions (still all details of basis construction
are worth a separate paper and cannot be covered here).

To use the SBases, one needs to create a start file or load it the same was as it has been done for
the FIRE algorithm. Now one has the following syntax:

\begin{itemize}

\item: \textit{Syntax:} {\tt BuildBasis[region,ordering]};

\textit{Description:} Tries to construct an $s$-basis
in the specified region with the given ordering.
The ordering should be an $n\times n$ non-degenerate positive-definite matrix.
The usage of region-bases (in case when the region is different from a sector)
is a special subject that is discussed in Section \ref{regions};
If the code succeeds in constructing
a basis, it prints the ``Evaluation successful'' message and a summary
on the expected master integrals.
It is worth noting that finding a proper ordering to construct a basis might be a difficult task
and requires a certain skill in complicated situations.

\textit{Example:} BuildBasis[\{1, 1, 1, 1\},\{\{1, 1, 1, 1\},\{1, 1, 1, 0\},\{1, 1, 0, 0\},\{1, 0, 0, 0\}\}];

\item: \textit{Syntax:} {\tt BuildBasis[sector]};

\textit{Description:} Try to use the random ordering mode to try and construct
a basis. The random mode is supported only for sectors;

\textit{Example:} BuildBasis[\{1, 1, 1, 1\}];

\item: \textit{Syntax:} {\tt Info[region]};

\textit{Description:} Provide the information on sectors
contained in the specified region;

\textit{Example:} Info[\{0, 0, 0, -1\}];

\item: \textit{Syntax:} {\tt Info[region,True]};

\textit{Description:} Same as {\tt Info[region]} but with
a listing of sectors;

\textit{Example:} Info[\{0, 0, 0, -1\},True];

\item: \textit{Syntax:} {\tt BuildAll[region]};

\textit{Description:} launches an automatic
attempt to build the $s$-bases in all sectors contained in the specified region.
The algorithm is bases on trying to find a single element and an ordering such
that this element produces proper expressions for all integrals in the given sector.
Hence this command might produce the $s$-bases only in the sectors
that have no master integrals. Anyway this might
be a good addition to the Laporta algorithm;

\textit{Example:} BuildAll[\{0, 0, 0, -1\}];

\item: \textit{Syntax:} {\tt SaveSBases[file\_without\_extension]};

\textit{Description:} saves the $s$-bases to a file;

\textit{Example:} {\tt SaveSBases["box"]} writes to {\tt box.sbases}.

\item \textit{Syntax:} {\tt F[$\{a_1,\ldots,a_n\}$]};

\textit{Description:} To verify how a basis works right after
constructing it, one can run the {\tt Burn[]} command
and {\tt F[$\{a_1,\ldots,a_n\}$]} without specifying the problem
number;

\textit{Example:} {\tt F[\{2,3,0,-4\}]};

\end{itemize}

\section{Examples}

\label{example}
Our first example is a massless on-shell box diagram (Fig.~\ref{box}) .

{\tt
$\;$\\
\indent Input 1: \\
\indent Get["FIRE\_3.0.0.m"];\\
\indent Get["IBP.m"];\\
\indent Internal = \{k\};\\
\indent External = \{p$_1$, p$_2$, p$_4$\};\\
\indent Propagators = \{-k$^2$, -(k + p$_1$)$^2$, -(k + p$_1$ + p$_2$)$^2$, -(k + p$_1$ + p$_2$ + p$_4$)$^2$\};\\
\indent PrepareIBP[];\\
\indent reps = \{p$_1^2 \rightarrow$0, p$_2^2 \rightarrow$0, p$_4^2 \rightarrow$0, p$_1$p$_2 \rightarrow$s/2, p$_2$p$_4 \rightarrow$ t/2, p$_1$p$_4 \rightarrow$-(s+t)/2\}\\
\indent startinglist = \{IBP[k, k], IBP[k, p$_1$], IBP[k, p$_2$], IBP[k, p$_4$]\}/.reps;\\
\indent RESTRICTIONS=\{\{-1,-1,0,0\},\{0,-1,-1,0\},\{0,0,-1,-1\},\{-1,0,0,-1\}\};\\
\indent SYMMETRIES = \{\{3, 2, 1, 4\}, \{1, 4, 3, 2\}, \{3, 4, 1, 2\}\};\\
\indent Prepare[];\\
\indent SaveStart["box"];\\
\indent Burn[];\\
\indent \\
\indent Output 1:\\
\indent FIRE, version 3.0.0\\
\indent UsingIBP: True\\
\indent UsingFermat: False\\
\indent Prepared\\
\indent Dimension set to 4\\
\indent Saving initial data
\\
}

Now the code is ready to work with the box diagram in the Laporta mode.
Simply run something like {\tt F[\{2, 2, 2, 2\}]} to get the answer.

To build the $s$-bases one has to load {\tt SBases\_3.0.0.m} and run the following commands:

{\tt
$\;$\\
\indent BuildBasis[\{1,1,1,1\},\{\{1,1,1,1\},\{1,1,1,0\},\{1,1,0,0\},\{1,0,0,0\}\}];\\
\indent BuildBasis[\{-1,1,1,1\},\{\{1,1,1,1\},\{1,1,1,0\},\{1,1,0,0\},\{1,0,0,0\}\}];\\
\indent BuildBasis[\{1,-1,1,1\},\{\{1,1,1,1\},\{1,1,1,0\},\{1,1,0,0\},\{1,0,0,0\}\}];\\
\indent BuildBasis[\{-1,1,-1,1\},\{\{1,1,1,1\},\{1,1,1,0\},\{1,1,0,0\},\{1,0,0,0\}\}];\\
\indent BuildBasis[\{1,-1,1,-1\},\{\{1,1,1,1\},\{1,1,1,0\},\{1,1,0,0\},\{1,0,0,0\}\}];
\\
}

Afterwards the evaluation will go faster than in the Laporta mode.

\begin{figure}
\begin{center}
\fcolorbox{white}{white}{
  \begin{picture}(158,150) (32,-30)
    \SetWidth{0.5}
    \SetColor{Black}
    \Line(105,75)(60,45)
    \Line(150,105)(105,75)
    \Line(105,15)(60,45)
    \Line(150,-15)(105,15)
    \Vertex(60,45){1.41}
    \Vertex(105,75){1.41}
    \Vertex(150,105){1.41}
    \Vertex(150,-15){1.41}
    \Vertex(105,15){1.41}
    \Line(105,75)(105,15)
    \Line(150,105)(150,-15)
    \ArrowLine(173,120)(150,105)
    \ArrowLine(173,-30)(150,-15)
    \ArrowLine(58,45)(32,45)
    \Text(160,-18)[lb]{\small{\Black{$q_2$}}}
    \Text(160,99)[lb]{\small{\Black{$q_1$}}}
    \Text(119,90)[lb]{\small{\Black{$1$}}}
    \Text(121,-11)[lb]{\small{\Black{$3$}}}
    \Text(76,18)[lb]{\small{\Black{$6$}}}
    \Text(76,62)[lb]{\small{\Black{$2$}}}
    \Text(109,45)[lb]{\small{\Black{$5$}}}
    \Text(154,45)[lb]{\small{\Black{$4$}}}
  \end{picture}
}
\end{center}
\caption{
\label{FF}
A diagram contributing to the 2-loop massless quark formfactor.
}
\end{figure}
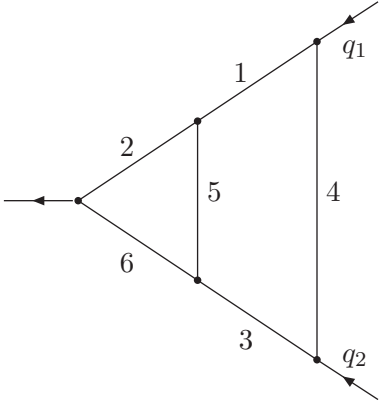

A more complicated example: a diagram contributing to the
two-loop massless quark formfactor (Fig.~\ref{FF}).

{\tt
$\;$\\
\indent Input 1: \\
\indent Get["SBases\_3.0.0.m"];\\
\indent Get["IBP.m"];\\
\indent Internal = \{k, l\};\\
\indent External = \{q$_1$, q$_2$\};\\
\indent Propagators = \{(l + q$_1$)$^2$, (k + q$_1$)$^2$, (l - q$_2$)$^2$, \\
\indent $\;\;\;\;\;\;\;\;\;\;$ l$^2$, (k - l)$^2$, (k - q$_2$)$^2$, k$^2$\};\\
\indent PrepareIBP[];\\
\indent reps=\{q$_1^2 \rightarrow$ 0, q$_2^2 \rightarrow$ 0, q$_1$ q$_2 \rightarrow$ -QQ/2\};\\
\indent startinglist = \{IBP[k, k], IBP[k, k-l], IBP[k, k+q$_1$], IBP[k, k-q$_2$], \\
\indent $\;\;\;\;\;\;\;\;\;\;$       IBP[l, l], IBP[l, l-k], IBP[l, l+q$_1$], IBP[l, l-q$_2$]\}/.reps;\\
\indent SYMMETRIES = \{\{3, 6, 1, 4, 5, 2, 7\}\};\\
\indent RESTRICTIONS = \{\{-1, -1, 0, 0, 0, 0, 0\}, \{0, 0, -1, 0, 0, -1, 0\},  \\
\indent $\;\;\;\;\;\;\;\;\;\;$ \{-1, 0, 0, 0, -1, 0, 0\}, \{0, -1, 0, 0, -1, 0, 0\},\\
\indent $\;\;\;\;\;\;\;\;\;\;$ \{0, 0, -1, 0, -1, 0, 0\}, \{0, 0, 0, 0, -1, -1, 0\},\\
\indent $\;\;\;\;\;\;\;\;\;\;$ \{0, -1, 0, 0, 0, -1, 0\}, \{-1, 0, -1, -1, 0, 0, 0\}\};\\
\indent Prepare[];\\
\indent \\
\indent Output 1:\\
\indent SBases, version 3.0.0\\
\indent UsingIBP: True\\
\indent UsingFermat: False\\
\indent Prepared\\
\indent Dimension set to 7
\\
}

In this example we demonstrate the automatic construction of $s$-bases.
First of all, let us list the sectors:

{\tt
$\;$\\
\indent
Input 2: \\
\indent Info[\{0, 0, 0, 0, 0, 0, -1\}]\\
\indent \\
\indent Output 2:\\
\indent Sectors in the  area :64\\
\indent Non-trivial  sectors  in  the  area :17\\
\indent Non-trivial  sectors  up  to  a  symmetry  in  the  area  :11\\
\indent Bases  built  in  0  sectors\\
\indent Rules  exist  in  0  sectors\\
\indent Nothing  in  11  sectors
\\
}

The last index is always non-positive for the reason that it
corresponds to an irreducible numerator. One can see
that we have provided enough information for the code
to avoid considering trivial sectors and to identify
symmetrical sectors. Now let us construct most $s$-bases automatically:

{\tt
$\;$\\
\indent
Input 3: \\
\indent BuildAll[\{0, 0, 0, 0, 0, 0, -1\}];
\\
}

We skip the output. Let us repeat the information command.
Now we will give it with the {\tt True} option forcing the
code to produce more output:

{\tt
$\;$\\
\indent Input 4: \\
\indent Info[\{0, 0, 0, 0, 0, 0, -1\}, True]\\
\indent \\
\indent Output 4:\\
\indent Sectors in the area :64\\
\indent Non-trivial sectors in the area :17\\
\indent Non-trivial sectors up to a symmetry in the area :11\\
\indent Bases built in 7 sectors\\
\indent Rules exist in 0 sectors\\
\indent Nothing in 4 sectors\\
\indent 2 additional minuses\\
\indent \{\{-1,1,-1,1,1,1,-1\}\}\\
\indent \{\{1,1,1,-1,-1,1,-1\}\}\\
\indent \{\{-1,1,1,1,1,-1,-1\},\{1,-1,-1,1,1,1,-1\}\}\\
\indent 3 additional minuses\\
\indent \{\{-1,1,1,-1,1,-1,-1\},\{1,-1,-1,-1,1,1,-1\}\}
\\
}

Each of the lines in the sector listing stands for a set of
symmetrical sectors, where one has no $s$-bases constructed.
The remaining sectors are likely to have master integrals.
One might try to construct the bases there manually.



\section{Advanced topic: usage of regions}

\label{regions}
The regions were introduced into the code to be able to perform integration
over loop momenta explicitly.
Consider a subset of indices $I$ and the
corresponding region $\nu=\{a_1,\ldots,a_n\}$ where $a_i$ is nonzero
if and only if $i\in I$. One tries to perform a reduction to masters
only considering the indices in $I$ as shift operators and the others
as coefficients. Afterwards the indices in $I$ can have only a finite number
of values (corresponding to master integrals of the subdiagram),
so a rule can be created mapping those integrals to a diagram
with less indices and loops.

This approach affects both the basis construction and the reduction.
However the most important difference is that the global ordering
has to be changed in a way. The reason is that the sectors inside
$\sg_\nu$ cannot be compared normally; instead, the \textit{active indices}
(the ones in $I$) should be compared as normal, and the \textit{passive indices}
(the remaining) can be changed in an arbitrary manner.

To make things even more complicated, we should keep in mind that there might
be different regions defined at the same time corresponding to
different subdiagrams. Therefore we had to change the definition of an ordering.
After fixing a problem we define the set of regions in this problem
(the syntax is {\tt SBasisM[pn]}). The order of regions in this list is important:
the regions going later are supposed to be lower. Moreover, for each point
$A=\{p_1,\ldots,p_n\}$ we consider the set of regions it lies in
and define \textit{the maximum region number} ($r_A$)
as the maximum position number for those regions inside the {\tt SBasisM[pn]}
(for a point that is contained in no regions we say that $r_A=0$).
Now we say that if $r_A<r_B$ then $B$ is lower than $A$ and vice versa.
The points with $r_A=0$ are compared in the standard way. Now let us consider
two points with equal nonzero maximal region numbers. First of all,
the active index values are taken and are compared in the standard way.
Next, if those coincide, we have to consider the passive indices,
which may be of any sign. Hence the sectors corresponding
to this subset of indices are considered and compared.
Finally if the points lie in the same subsector, they are compared
with the use of the ordering.

This approach has the following implications: first, we have
to define a unique ordering for each region, and this ordering has to be specific:
the active indices have to be ``more important'' than the passive ones.
Hence the ordering matrix should be of a ``block shape'':
the first $k=number\;of\;active\;indices$ rows can have non-zero numbers only in the columns
corresponding to the active indices, the following rows
can have non-zero numbers only in the columns
corresponding to the passive ones.

Now, having chosen such an ordering, one can try to construct a region basis with the {\tt BuildBasis[region,ordering]} command.

The second implication is that one should not be able to reduce a point inside a region
to a point not lying in a region with a greater number.
This could happen if some of the active indices turned negative (from the positive
value inside the region) and there were no region there.
Hence we require that the points of that type
should either correspond to integrals vanishing by the boundary conditions,
or lie in a ``parallel region'' (with the same set of zeros in the definition)
having a greater number.

\section{Perspectives}

The beta-version of FIRE has been applied to a family of three-loop Feynman integrals necessary for the analysis of decoupling of $c$-quark loops in $b$-quark HQET~\cite{GSS}.
Furthermore, the algorithm itself has been used to calculate the fermionic contribution to the three-loop static quark potential \cite{tlp}.
FIRE has also been used to crosscheck some of the integrals in a paper~\cite{MMM} where four-loop vacuum integrals have been evaluated.

The current version of the FIRE algorithm allows to perform difficult calculations involving millions of integrals.
One of the possible improvements in the future is to reproduce some parts of the algorithm in C to increase the productivity and enable the parallelization.

\section{Acknowledgements}

I am grateful to V.A.~Smirnov for giving me this brilliant possibility
to work on the reduction of Feynman integrals, --- without a person
knowing physics well and applying the code constantly on real examples
(and finding numerous bugs) I would never be able to result in a
reasonably powerful code.

Special thanks to M.~Steinhauser for the collaboration on evaluating
the three-loop static quark potential, for
testing the code for a long time and for reading the drafts of my paper.

I would like to thank M.~Tentioukov for consulting me on the Laporta algorithm,
as well as on the help on coding in C for Linux and the gateToFermat library.

I would also like to thank K.G.~Chetyrkin, A. Maier, P. Maierhoefer, P. Marqaurd and A.G.~Grozin
for our joint work on different physical problems resulting in
additional functionality of FIRE covering the specifics of those problems.

And of course I would like to thank the Institute for Theoretical Particle Physics of the University
of Karlsruhe for the hospitality it offers and the powerful machines
required for running the code.

The work has been supported by RFBR grant 08-02-01451-a and DFG through SBF/TR 9.

\end{document}